\documentclass{PoS}
\usepackage[authoryear,square]{natbib}
\bibpunct{(}{)}{;}{a}{}{,}
\let\ifpdf\relax

\title{SKA - EoR correlations and cross-correlations: \\kSZ, radio galaxies, and NIR background}

\ShortTitle{SKA-EoR correlations and cross-correlations}

\author{\speaker{Vibor Jeli\'{c}$^{1,2}$},
	Benedetta Ciardi$^3$,
	Elizabeth Fernandez$^1$,
	Hiroyuki Tashiro$^4$,
	Dijana Vrbanec$^3$
	\\
        $^1$Kapteyn Astronomical Institute, University of Groningen, PO Box 800, 9700 AV Groningen, the Netherlands; \\
        $^2$ASTRON - the Netherlands Institute for Radio Astronomy, PO Box 2, 7990 AA Dwingeloo, the Netherlands; \\
       	$^3$Max-Planck Institute for Astrophysics, Karl-Schwarzschild-Strasse 1, D-85748 Garching bei M\"unchen, Germany; \\ 
        $^4$Department of Physics and Astrophysics, Nagoya University, Furocho, Chikusaku, Nagoya, 464-8602 Japan\\
	\\ 
        E-mail: \email{vjelic@astro.rug.nl}}
	
\abstract{The Universe's Cosmic Dawn (CD) and Epoch of Reionization (EoR) can be studied using a number of observational probes that provide complementary or corroborating information. Each of these probes suffers from its own systematic and statistical uncertainties. It is therefore useful to consider the mutual information that these data sets contain. In this paper, we discuss a potential of cross-correlations between the SKA cosmological 21~cm data with: (i) the kinetic Sunyaev-Zel'dovich (kSZ) effect in the CMB data; (ii) the galaxy surveys; and (iii) near infrared (NIR) backgrounds.}

\FullConference{
Advancing Astrophysics with the Square Kilometre Array\\
June 8-13, 2014\\
Giardini Naxos, Italy}

\begin{document}
\section{Introduction}
Experiments designed to measure the redshifted 21 cm line from the Cosmic Dawn (CD) and Epoch of Reionization (EoR) are challenged by the strong astrophysical foreground contamination, ionospheric distortions, radio frequency interference and complex instrumental response. In order to reliably detect the cosmological signal from the observed data, it is essential to understand in detail all aspects of the experiment. For example, the cosmological signal has some characteristics which differentiates it from the foregrounds and noise. By use of proper statistics, it is possible to remove these components 
to extract signatures of reionization \citep[see][]{chapmanPoS}. 

To alleviate some of the problems associated with the observations of the weak cosmological signal, several cross-correlation analyses with observations in other frequency windows have been proposed. In doing this, the noise/systematics in two observations of different frequencies and strategies may cancel out. Even if SKA may have a high enough sensitivity not to need cross-correlation techniques in order to detect the signal, cross-correlating with other probes will improve our understanding of the process of reionization \citep[for an overview][]{mellema13}.

In this paper we present potential of cross-correlation studies between the SKA cosmological 21 cm data, which reflects the neutral hydrogen content of the Universe as a function of redshift, with: (i) the kinetic Sunyaev-Zel'dovich (kSZ) effect in the Cosmic Microwave Background (CMB) data, produced by the scattering of CMB photons off free electrons produced during the reionization process (Sec.~\ref{sec:kSZ}); (ii)  galaxy surveys (Sec.~\ref{sec:gal}); and (iii) the near infrared (NIR) backgrounds that reflect the primordial star formation (Sec.~\ref{sec:NIRB}). 

Throughout this study we consider three SKA configurations: early science SKA1-LOW, SKA1-LOW and SKA2-LOW. The SKA1-LOW \citep{SKA1} is
our standard SKA configuration for the noise simulations, using the OSKAR simulator\footnote{http://www.oerc.ox.ac.uk/~ska/oskar}. We assume full correlation between all 866 core stations of SKA1-LOW, where the maximum baseline length is 5.29 km. Baseline coordinates are generated for a 12-hour synthesis observation, with a 5-minute sampling interval. The noise is then re-normalised to reflect 1000 hours of integration. For early science SKA1-LOW we assume that the sensitivity is halved. For SKA2-LOW we assume that the sensitivity is quadrupled.

\section{The Cross-Correlation with the kSZ}\label{sec:kSZ}
One of the leading sources of secondary anisotropies in the CMB is due to the scattering of CMB photons off free electrons \citep{zeldovich69}. The effect of anisotropies 
when induced by thermal motions of free electrons is called the thermal Sunyaev-Zel'dovich effect (tSZ) and when due to bulk motion of free electrons is called the kinetic Sunyaev-Zel'dovich effect (kSZ). The latter is far more dominant during reionization \citep[for a review of secondary CMB anisotropies see, e.g.][]{aghanim08}.

The kSZ effect from a homogeneously ionized medium, i.e., with ionized fraction only as a function of redshift, has been studied both analytically and numerically by a number of authors; the linear regime of this effect was first calculated by \citet{sunyaev70} and subsequently revisited by \citet{ostriker86} and \citet{vishniac87} -- hence also referred to as the Ostriker-Vishniac (OV) effect. In recent years various groups have calculated this effect in its non-linear regime using semi-analytical models and numerical simulations
\citep{gnedin01, santos03, zhang04}. These studies show that the contributions from non-linear effects are only important at small angular scales ($l>1000$), while the OV effect dominates at larger angular scales.

The kSZ effect from patchy reionization was first estimated using simplified semi-analytical models by \citet{santos03}, who concluded that it dominates over that of a homogeneously ionized medium. More detailed modeling of the effect of patchy reionization were subsequently performed using numerical simulations \citep{salvaterra05, iliev07} and semi-analytical models \citep{mcquinn05, zahn05,mesinger12}. \citet{dore07} used numerical simulations to derive the expected CMB polarization signals
due to EoR patchiness. The CMB bolometric arrays Atacama Cosmology Telescope \citep[ACT, ][]{fowler10} and South Pole Telescope \citep[SPT, ][]{shirokoff11} are currently being used to measure the CMB anisotropies at the scales relevant to reionization ($3000<\ell<8000$). The SPT results are starting to put limits on the duration of reionization \citep{zahn12}.

\begin{figure}
\centering \includegraphics[width=.51\textwidth]{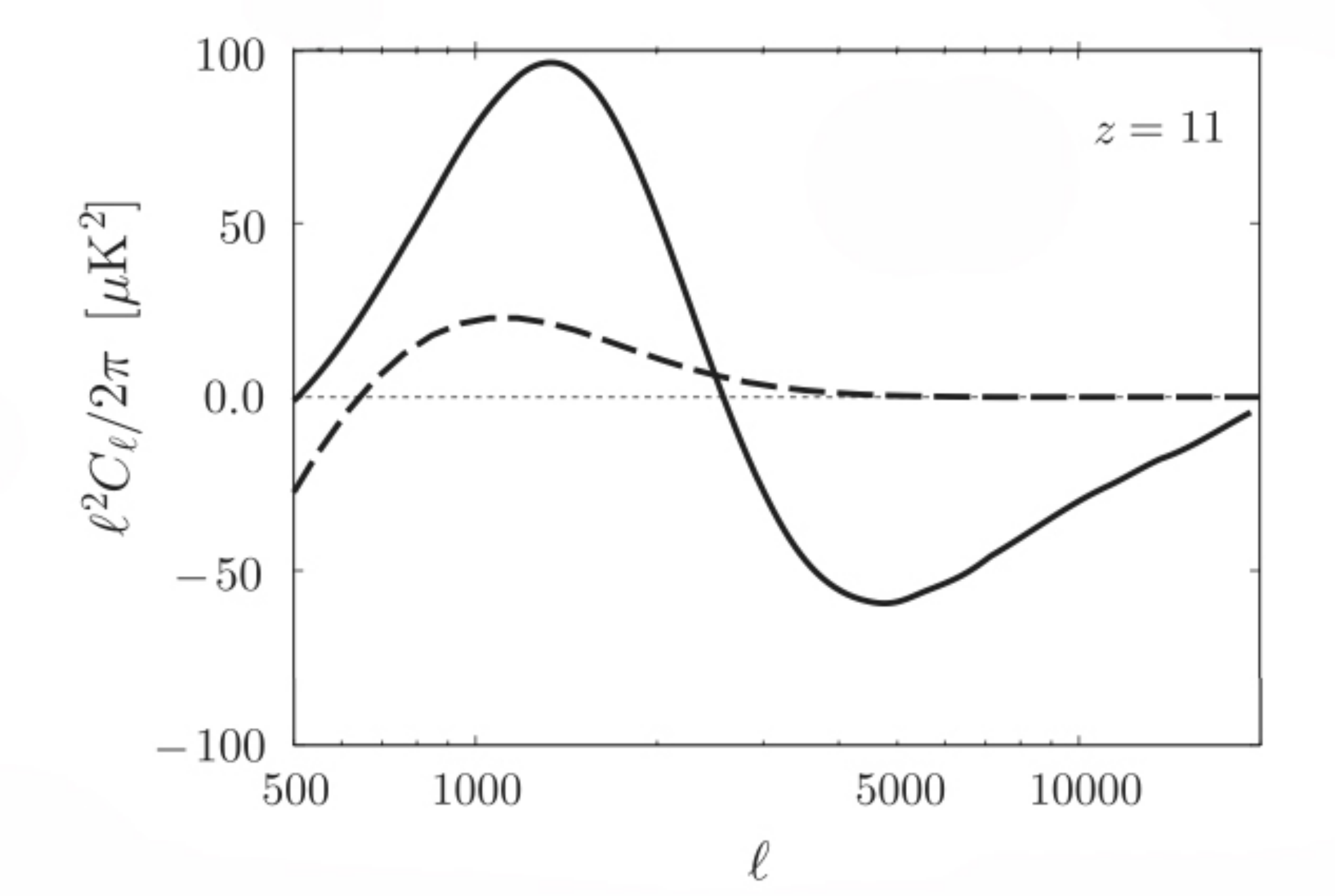}
\caption{An example of the cross-power spectrum of the kSZ and the cosmological 21cm signal at $z=11$.  The solid line is for a `patchy' reionization history, while the dashed line is for  a `homogenous' history.  \citep{tashiro11}}
\label{fig:kSZEoRcorr1}
\end{figure}

\begin{figure}
\centering \includegraphics[width=.49\textwidth]{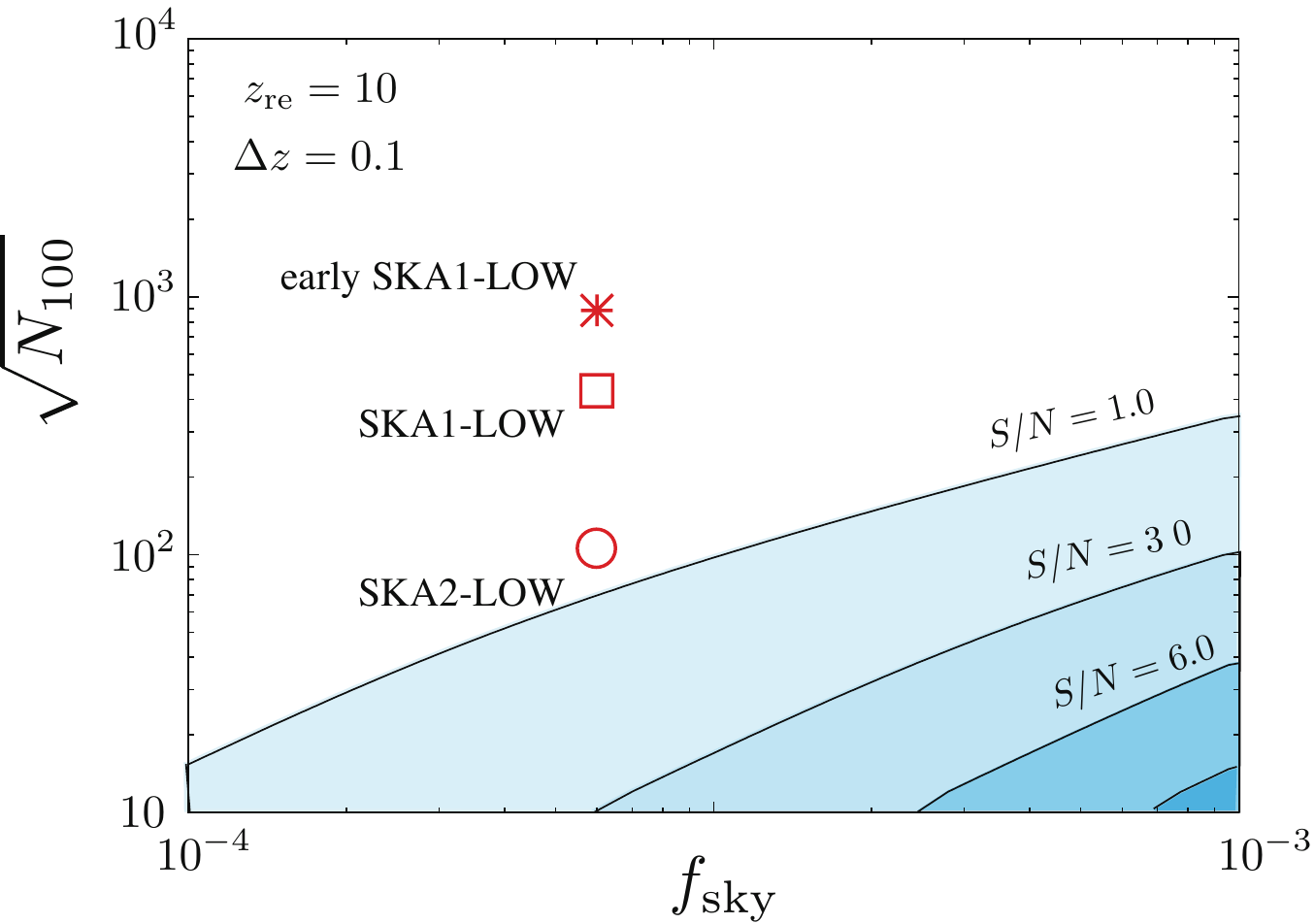}
\centering \includegraphics[width=.49\textwidth]{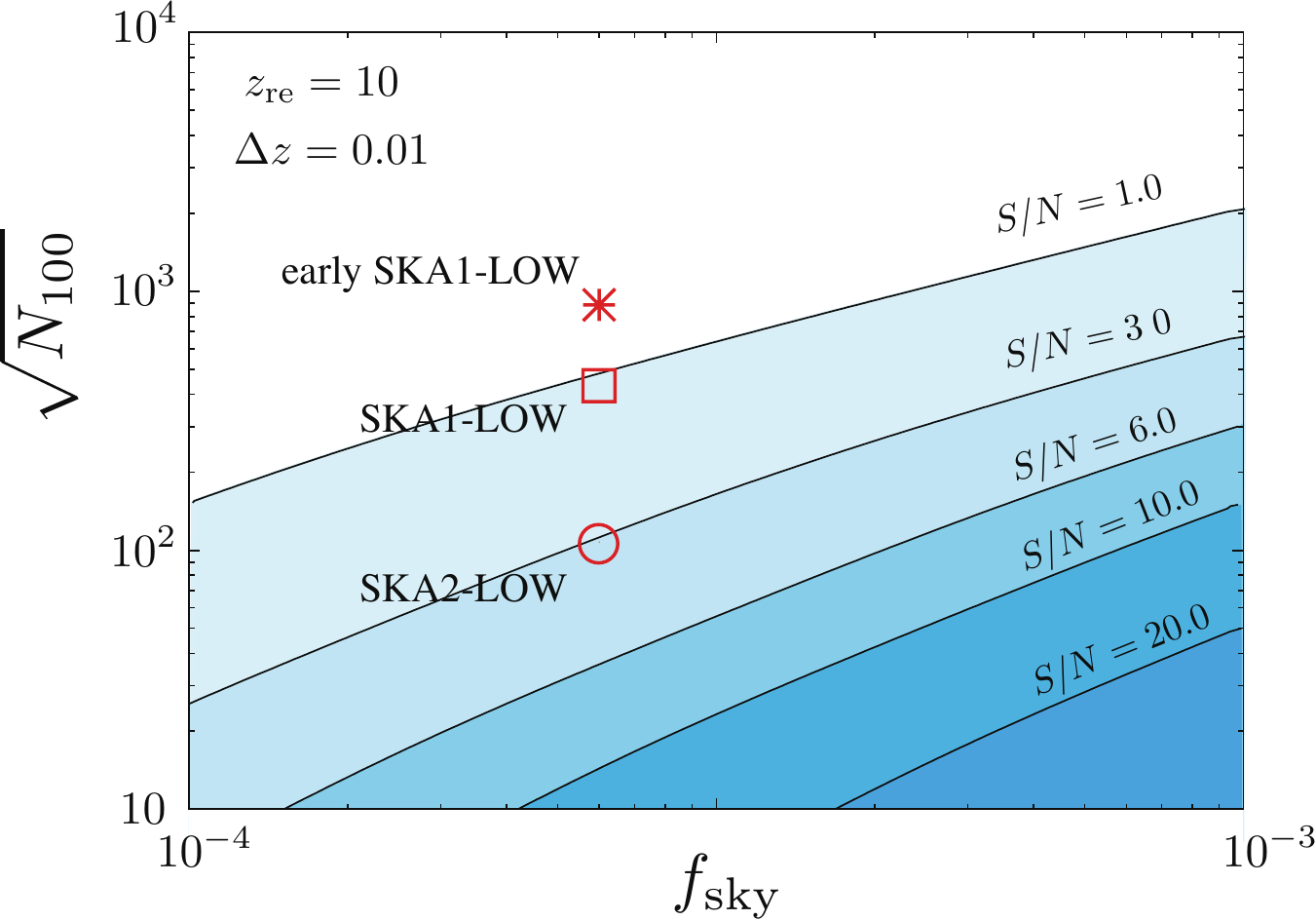}
\caption{S/N of the 21cm cross-correlation with the kSZ for two different reionization models, defined with the redshift at which the ionized fraction equals 0.5 ($z_{\rm re}$) and the reionization duration ($\Delta z$). The S/N is given as a function of the sky fraction and the normalised noise power spectrum. We assume 1000h of integration time with early SKA1-LOW, SKA1-LOW and SKA2-LOW and a field of view of $25~{\rm deg^2}$. For the CMB data we assume Planck sensitivity. \citep[based on][]{tashiro10}}
\label{fig:kSZEoRcor2}
\end{figure}

Cross-correlation between the cosmological 21cm signal, as measured with SKA, and the secondary CMB anisotropies provide a potentially useful statistic.   The cross-correlation has the advantage that the measured statistic is less sensitive to contaminants such as the foregrounds, systematics and noise in comparison to ``auto-correlation'' studies.

Analytical cross-correlation studies between the CMB temperature anisotropies and the EoR signal on large scales ($l\sim100$) were carried out by \citet{alvarez06, adshead08, lee09} and on small scales ($l>1000$) by \citet{cooray04,salvaterra05,slosar07,tashiro08,tashiro10,tashiro11}.  Cross-correlation between the E- and
B-modes of CMB polarization with the redshifted 21cm signal was done by \citet{tashiro08,dvorkin09}. Numerical studies of the cross-correlation were carried out by \citet{salvaterra05, jelic10}.

These studies showed that the kSZ and the redshifted 21cm signal: (i) anti-correlate on the scales corresponding to the typical size of ionized bubbles; and (ii) correlate on the larger scales, where the patchiness of the ionization bubbles are averaged out (see Fig.~\ref{fig:kSZEoRcorr1}). The significance of the anti-correlation signal depends on the reionization scenario \citep{salvaterra05, jelic10,tashiro11}.

The cross-correlation signal turns out to be difficult to detect (see Fig.~\ref{fig:kSZEoRcor2}). We might be able to make detection only with SKA2-LOW, assuming a very radical reionization duration of $\Delta z=0.01$. If duration of reionization is longer, detection is not possible for all three SKA configurations. However, the kSZ signal induced during the EoR could possibly be detected in the power spectra of the CMB and used to place some additional constraints on this epoch in the history of our Universe.

\section{The Cross-Correlation with Galaxy Surveys}\label{sec:gal}
Following \cite{lidz09}, one can define the cross power spectrum between the 21cm emission and the galaxies as:

\begin{equation}
\begin{array}{lll}\Delta^2_{\rm 21,gal}(k) &=& \tilde{\Delta}^2_{\rm 21,gal}(k)/\delta T_{b0} \\
& = &  x_{\rm \textsc{hi}} \left[ \Delta^2_{x_\mathrm{HI},{\rm
       gal}}(k) + \Delta^2_{\rho,{\rm gal}}(k) \right.\\
& & \left. + \Delta^2_{\rho x_\mathrm{HI},{\rm gal}}(k)\right],
\end{array}
\label{eq-cps}
\end{equation}
where $\delta T_{b0}$ is the 21cm brightness temperature relative to the CMB for neutral gas at the mean density of the universe, $x_{\rm HI}$ is the neutral fraction and $\Delta^2_{a,b}(k)$ is the dimensionless cross power spectrum between fields $a$ and $b$.  In order to construct the cross power spectrum, one therefore requires three fields, the density field ($\rho$), the neutral hydrogen field ($x_{\rm HI}$), and the galaxy field (gal), which can be obtained via numerical simulations of galaxy formation and the reionization
process.

\begin{figure}[t!]
\begin{center}
\includegraphics[width=1.\textwidth]{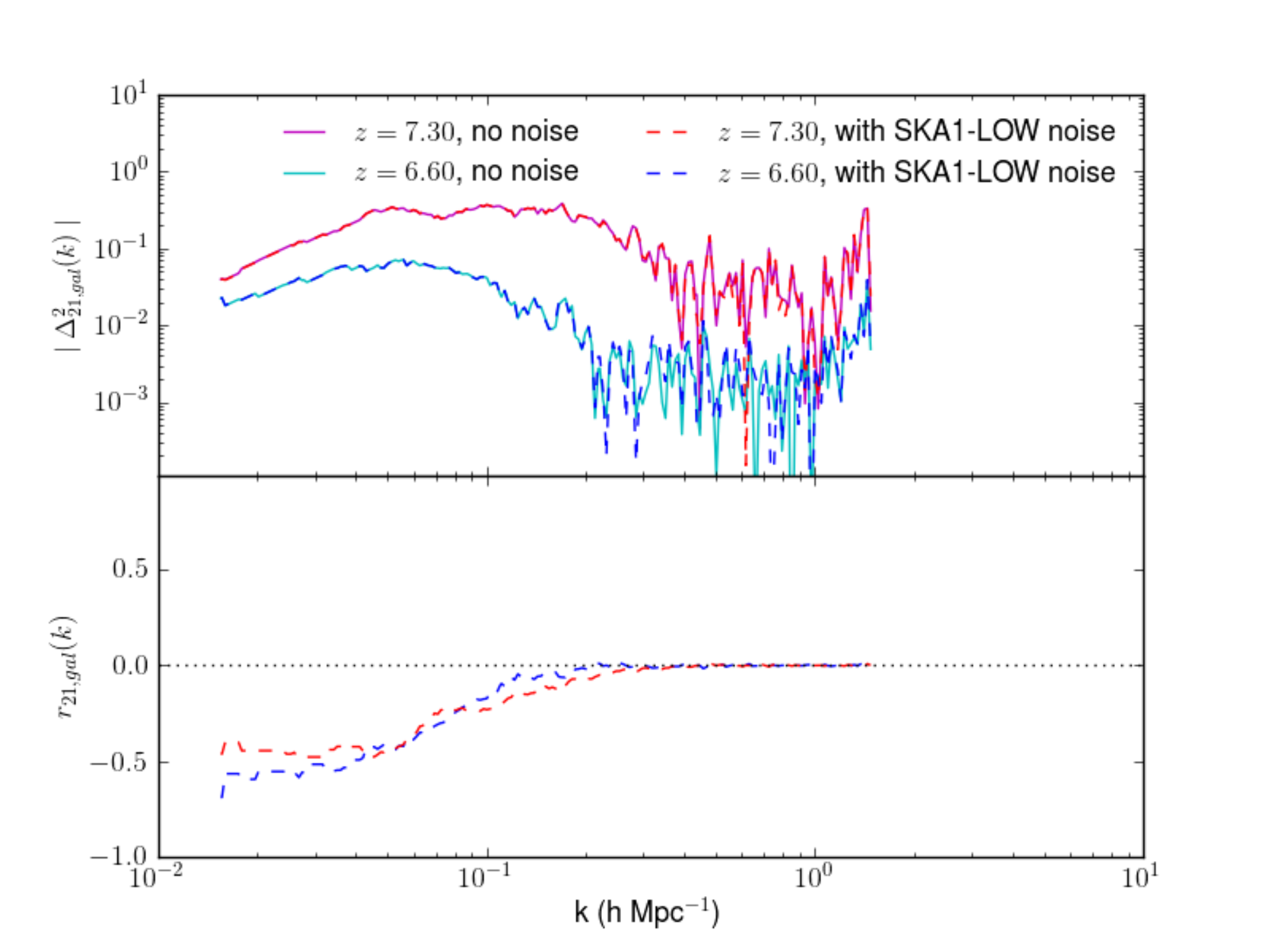}
\caption{The circularly averaged, unnormalized 2D 21cm --
 galaxy cross power spectrum ($\tilde{\Delta}^2_{\rm 21,gal}(k)$;
 and correlation coefficient for two
 redshifts. In the analysis we followed \citet{wiersma13}.}
\label{fig:gal21cm}
\end{center}
\end{figure}

It is found that the 21cm emission is initially correlated with galaxies on large scales, anti-correlated on intermediate scales, and uncorrelated on small scales. This picture quickly changes as reionization proceeds and the two fields become anti-correlated on large scales \citep{lidz09,wiersma13}. These (anti-) correlations can be a powerful tool in indicating the topology of reionization and should form important diagnostic tools for SKA observations.

If the effect of observing and selecting real galaxies is taken into account, the result depends on the observational campaign considered. For example, for a drop-out technique (as in observations of Lyman Break Galaxies), the normalization of the cross power spectrum seems to be the most powerful tool for probing reionization. In particular, it is quite sensitive to the ionized fraction as different reionization histories yield similar cross power spectra for a fixed ionized fraction.  When instead a more precise measurement of the galaxy redshifts is available (as in Ly-$\alpha$ Emitters surveys), which provides the three-dimensional position of the galaxy, much more information about the nature of reionization can
be extracted, since the shape and the normalization of the cross power spectrum provide useful information. In addition, the observability of the Ly-$\alpha$ line from these galaxies is affected by neutral patches in the IGM and thus Ly-$\alpha$ Emitters surveys are particularly useful for EoR studies \citep{mcquinn07, jensen14}.

Figure \ref{fig:gal21cm} shows the 21cm - Ly-$\alpha$ Emitters cross power spectrum (with and without SKA noise) and correlation coefficient for two redshifts. Here the noise assigned to the 21cm survey is the one of the {SKA} telescope with SKA1-LOW configuration and noise for 1000 hours of integration time, while the Ly-$\alpha$ Emitters survey has the same characteristics of the one described in \citet{jensen14,ouchi10} with the {Subaru} telescope. As we can see, the 21cm - Ly-$\alpha$ Emitters cross power spectrum is not very sensitive to noise. Therefore, early SKA1-LOW and SKA2-LOW will show very similar results. The effect of neutral patches in the IGM on the observability of these Ly-$\alpha$ Emitters is not included here.

\section{The Cross-Correlation with Near Infrared Backgrounds}\label{sec:NIRB}
Understanding star formation at high redshifts is fundamentally linked to our understanding of reionization history.  In order for reionization to occur, a plentiful source of ionizing photons was needed.  Stars are a likely candidate to produce these photons.  Therefore, any attempt to understand reionization must be paired with an understanding of primordial star formation.

However, current observational constraints suggest that in order for the Universe to be reionized, a plentiful number of small galaxies beneath the detection limit of current surveys are most likely needed.  Because they are so faint, it is difficult to obtain information about these galaxies directly.  Instead, these galaxies can be observed using indirect means.  For example, the cumulative light from these galaxies would be redshifted to the infrared, and therefore, any background in the infrared may provide clues to the nature of these galaxies.  Many have suggested that the remnant light from these galaxies is indeed present in the excess emission in the Near Infrared Background (NIRB), and if this is true, looking at the mean intensity and the power spectrum of the NIRB can give information about these early stellar populations  \citep[e.g.,][]{santos02, magliocchetti03, salvaterra03, cooray04, kashlinsky05, fernandez06, fernandez10, cooray12, fernandez12, kashlinsky12, fernandez13, yue13}. 

On the other hand, 21cm emission originates from areas of neutral hydrogen, so it corresponds to regions that have not seen plentiful star formation and hence have not yet been ionized.  Therefore, we should expect that regions that are bright in 21cm background emission should not be bright in the infrared.  Conversely, regions that are bright in the infrared have plenty of star formation and therefore should be dim in the 21cm background.  Because of this, these two observations should be anti-correlated.  

In \citet{fernandez14}, we predict the observational properties of both high redshift galaxies (which would be observable in the infrared) and the neutral regions that have yet to be ionized by galaxies (observable as part of the 21cm background).  In order to do this, we combined large scale N-body simulations \citep{iliev14}, which included radiative transfer, with analytical models of the luminosities of the halos \citep{fernandez06}. The brightness temperature of the 21cm background was also computed from the simulation, which depends on the density and neutral fraction of the IGM around the galaxies.   

In \citet{fernandez14}, we then produced simulated sky maps for both the 21cm emission and the emission in the infrared.  These maps were created by combining the emission models and the simulation output to create luminosity cubes.  These cubes were then randomly rotated and stacked along the line of sight to create a three dimensional cuboid.  This was projected onto two dimensions to create a simulated sky map in both the radio and the infrared.  In  \citet{fernandez14}, we created sky maps specific for the LOFAR instrument; here, we present maps tailored for three SKA configurations: early science SKA1-LOW, SKA1-LOW and SKA2-LOW (assuming 1000h of integration). 

For the radio map, since the 21cm background is line emission, each frequency corresponds to a different redshift, and observations can be adjusted to correspond to any arbitrary redshift range.  On the other hand, since the emission in the infrared has a considerable amount of continuum, it is very difficult to extract redshift information for observations at any given wavelength.  Therefore, the map in the infrared is created using the integrated light from a redshift of 6 to 30, or the entire simulation volume.  

\begin{figure}
\centering \noindent
\includegraphics[width=0.48\textwidth]{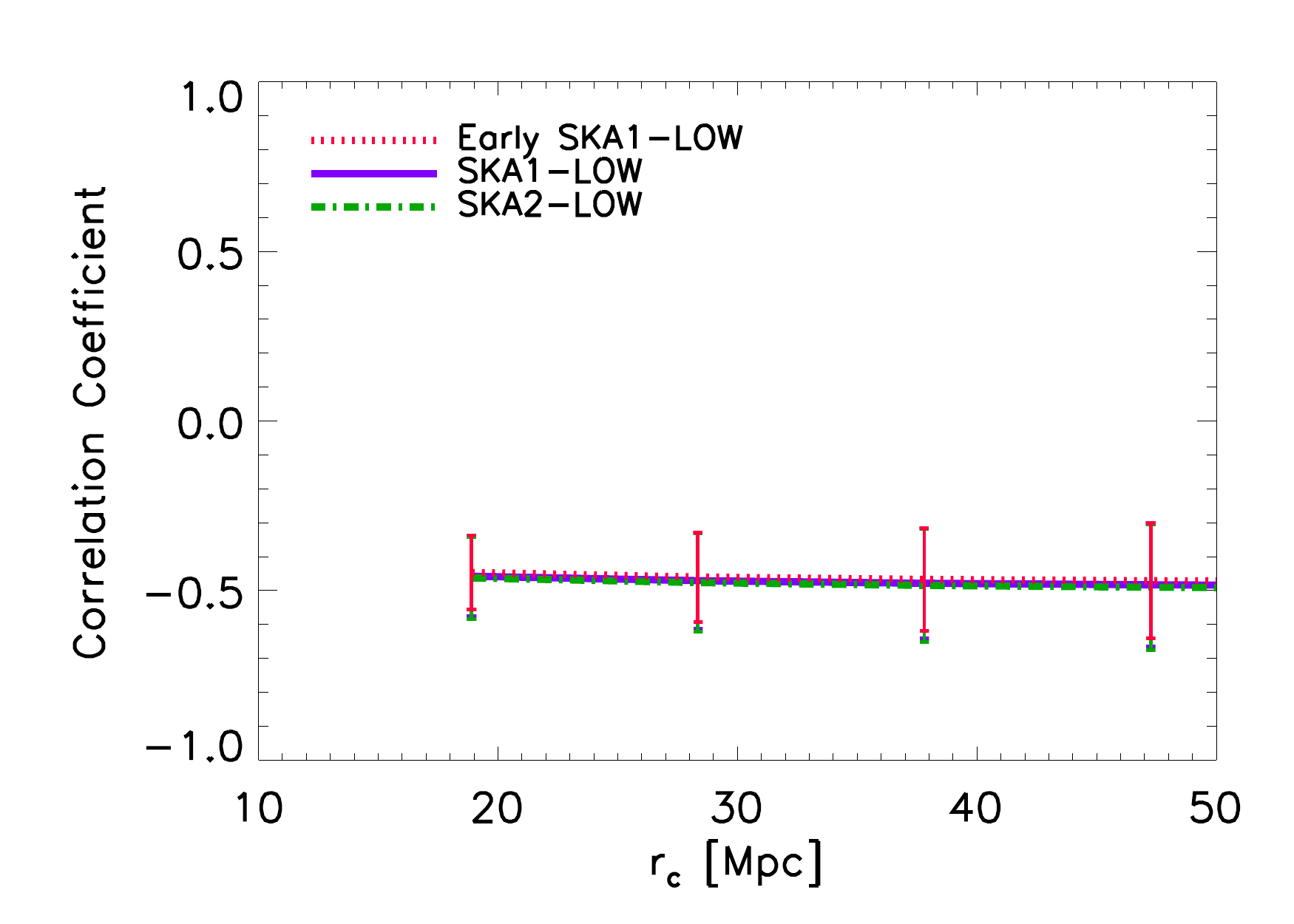}
\includegraphics[width=0.48\textwidth]{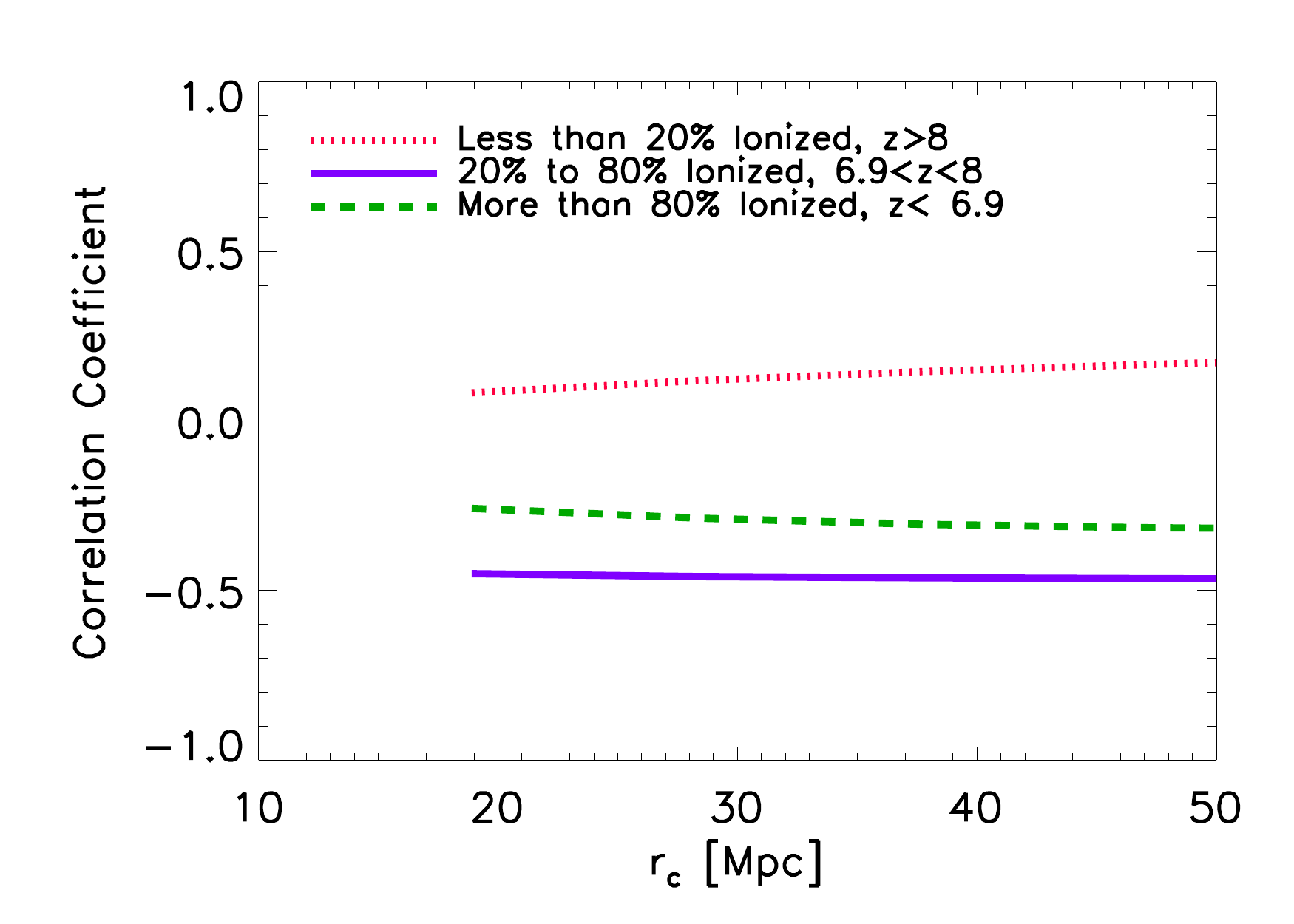}
\caption{Left: The cross correlation of the entire NIRB against the 21cm background, assuming three SKA configurations with 1000h of integration time. The results do not depend heavily on noise levels, indicating that errors from unknowns in observation dominate over errors introduced from the noise. 
Right: The cross correlation at various redshifts: $6.4<z<7$, $7<z<8$, and $8<z<11$, assuming SKA1-LOW configuration.}
\label{fig:IRcorr}
\end{figure}

These two maps in the infrared and in the radio can then be cross-correlated using the Pearson correlation coefficient:
\begin{equation}
\rho_{21 {\rm{cm}}, {\rm{NIRB}}} = \frac{{\rm{cov}}({(\delta T_{\rm{b}}),{I_{NIRB}}})}{\sigma_{\delta T_{\rm{b}}}\sigma_{I_{NIRB}}},
\end{equation}
where $I_{NIRB}$ is the intensity in the infrared and $\delta T_{\rm{b}}$ is the brightness temperature of the 21cm line.  The results of this cross-correlation is shown in left panel of Fig.~\ref{fig:IRcorr}, which illustrates the cross-correlation between the integrated light from $6<z<30$ in the near-infrared against the entire 21cm map from $6.4<z<11$, assuming three SKA configurations.  Error bars are generated from cross-correlating the 21cm emission with 1000 randomized maps in the infrared.  The cross-correlation coefficient shows that these two maps are strongly anti-correlated, corresponding to the fact that the two emission maps are sensitive to emission from mutually exclusive areas.  In addition, there is not a significant difference in result between Early SKA1-LOW, SKA1-LOW, and SKA2-LOW, indicating that the errors introduced from the noise is sub-dominant.  

In addition, since the 21cm background is line emission, we can cross-correlate only specific slices of the 21cm background with the entire NIRB.  Results of this for SKA1-LOW configuration are shown in the right panel of Fig.~\ref{fig:IRcorr}. The correlation coefficient is computed using radio maps for early ($z>8$), mid ($6.9<z<8$) and late ($z<6.9$) stages of reionization against the entire NIRB.  The correlation coefficient is the most negative during mid stages of reionization.  At early stages of reionization, ionized bubbles around the host galaxies are too small, perhaps smaller than the smoothing length, and therefore no correlation is seen between the two maps.  At late stages of reionization, most of the structure in the 21cm maps has disappeared.  
\section{Conclusions}
In the view of presented results, cross-correlations with other probes will improve our understanding of the process of reionization. The cross-correlation with the galaxy surveys and NIR backgrounds will specifically help in answering the question which types of galaxies are mostly responsible for reionization. There is not a significant difference in results between Early SKA1-LOW, SKA1-LOW, and SKA2-LOW, indicating that the errors introduced from the noise is sub-dominant. Unfortunately,  the cross-correlation signal in the case of the CMB turns out to be difficult to detect. There is only a possibility of detection in the case of a very short reionization scenario and using SKA2-LOW.

\bibliographystyle{apj}
\bibliography{reflist}

\begin{thebibliography}{48}
\expandafter\ifx\csname natexlab\endcsname\relax\def\natexlab#1{#1}\fi

\bibitem[{{Adshead} \& {Furlanetto}(2008)}]{adshead08}
{Adshead}, P.~J. \& {Furlanetto}, S.~R. 2008, \mnras, 384, 291

\bibitem[{{Aghanim} {et~al.}(2008){Aghanim}, {Majumdar}, \& {Silk}}]{aghanim08}
{Aghanim}, N., {Majumdar}, S., \& {Silk}, J. 2008, Reports on Progress in
  Physics, 71, 066902

\bibitem[{{Alvarez} {et~al.}(2006){Alvarez}, {Komatsu}, {Dor{\'e}}, \&
  {Shapiro}}]{alvarez06}
{Alvarez}, M.~A., {Komatsu}, E., {Dor{\'e}}, O., \& {Shapiro}, P.~R. 2006,
  \apj, 647, 840

\bibitem[{{Chapman} {et~al.}(2014){Chapman}, {Bonaldi}, {Harker}, {Jeli\'c},
  {Abdalla}, {Bernardi}, {Bobin}, {Dulwich}, {Mort}, {Santos}, \&
  {Starckh}}]{chapmanPoS}
{Chapman}, E., {Bonaldi}, A., {Harker}, G., {Jeli\'c}, V., {Abdalla}, F.,
  {Bernardi}, G., {Bobin}, J., {Dulwich}, F., {Mort}, B., {Santos}, M., \&
  {Starckh}, J.-L. 2014, PoS(AASKA14)

\bibitem[{{Cooray}(2004)}]{cooray04}
{Cooray}, A. 2004, \prd, 70, 063509

\bibitem[{{Cooray} {et~al.}(2012){Cooray}, {Gong}, {Smidt}, \&
  {Santos}}]{cooray12}
{Cooray}, A., {Gong}, Y., {Smidt}, J., \& {Santos}, M.~G. 2012, \apj, 756, 92

\bibitem[{{Dewdney} {et~al.}(2013){Dewdney}, {Turner}, {Millenaar}, {McCool},
  {Lazio}, \& {Cornwell}}]{SKA1}
{Dewdney}, P., {Turner}, W., {Millenaar}, R., {McCool}, R., {Lazio}, J., \&
  {Cornwell}, T. 2013, SKA-TEL-SKO-DD-001

\bibitem[{{Dor{\'e}} {et~al.}(2007){Dor{\'e}}, {Holder}, {Alvarez}, {Iliev},
  {Mellema}, {Pen}, \& {Shapiro}}]{dore07}
{Dor{\'e}}, O., {Holder}, G., {Alvarez}, M., {Iliev}, I.~T., {Mellema}, G.,
  {Pen}, U.-L., \& {Shapiro}, P.~R. 2007, \prd, 76, 043002

\bibitem[{{Dvorkin} {et~al.}(2009){Dvorkin}, {Hu}, \& {Smith}}]{dvorkin09}
{Dvorkin}, C., {Hu}, W., \& {Smith}, K.~M. 2009, ArXiv e-prints

\bibitem[{{Fernandez} {et~al.}(2012){Fernandez}, {Iliev}, {Komatsu}, \&
  {Shapiro}}]{fernandez12}
{Fernandez}, E.~R., {Iliev}, I.~T., {Komatsu}, E., \& {Shapiro}, P.~R. 2012,
  \apj, 750, 20

\bibitem[{{Fernandez} \& {Komatsu}(2006)}]{fernandez06}
{Fernandez}, E.~R. \& {Komatsu}, E. 2006, \apj, 646, 703

\bibitem[{{Fernandez} {et~al.}(2010){Fernandez}, {Komatsu}, {Iliev}, \&
  {Shapiro}}]{fernandez10}
{Fernandez}, E.~R., {Komatsu}, E., {Iliev}, I.~T., \& {Shapiro}, P.~R. 2010,
  \apj, 710, 1089

\bibitem[{{Fernandez} \& {Zaroubi}(2013)}]{fernandez13}
{Fernandez}, E.~R. \& {Zaroubi}, S. 2013, \mnras, 433, 2047

\bibitem[{{Fernandez} {et~al.}(2014){Fernandez}, {Zaroubi}, {Iliev}, {Mellema},
  \& {Jeli{\'c}}}]{fernandez14}
{Fernandez}, E.~R., {Zaroubi}, S., {Iliev}, I.~T., {Mellema}, G., \&
  {Jeli{\'c}}, V. 2014, \mnras, 440, 298

\bibitem[{{Fowler} {et~al.}(2010){Fowler}, {Acquaviva}, {Ade}, {Aguirre},
  {Amiri}, {Appel}, {Barrientos}, {Battistelli}, {Bond}, {Brown}, {Burger},
  {Chervenak}, {Das}, {Devlin}, {Dicker}, {Doriese}, {Dunkley}, {D{\"u}nner},
  {Essinger-Hileman}, {Fisher}, {Hajian}, {Halpern}, {Hasselfield},
  {Hern{\'a}ndez-Monteagudo}, {Hilton}, {Hilton}, {Hincks}, {Hlozek},
  {Huffenberger}, {Hughes}, {Hughes}, {Infante}, {Irwin}, {Jimenez}, {Juin},
  {Kaul}, {Klein}, {Kosowsky}, {Lau}, {Limon}, {Lin}, {Lupton}, {Marriage},
  {Marsden}, {Martocci}, {Mauskopf}, {Menanteau}, {Moodley}, {Moseley},
  {Netterfield}, {Niemack}, {Nolta}, {Page}, {Parker}, {Partridge}, {Quintana},
  {Reid}, {Sehgal}, {Sievers}, {Spergel}, {Staggs}, {Swetz}, {Switzer},
  {Thornton}, {Trac}, {Tucker}, {Verde}, {Warne}, {Wilson}, {Wollack}, \&
  {Zhao}}]{fowler10}
{Fowler}, J.~W., {Acquaviva}, V., {Ade}, P.~A.~R., {Aguirre}, P., {Amiri}, M.,
  {Appel}, J.~W., {Barrientos}, L.~F., {Battistelli}, E.~S., {Bond}, J.~R.,
  {Brown}, B., {Burger}, B., {Chervenak}, J., {Das}, S., {Devlin}, M.~J.,
  {Dicker}, S.~R., {Doriese}, W.~B., {Dunkley}, J., {D{\"u}nner}, R.,
  {Essinger-Hileman}, T., {Fisher}, R.~P., {Hajian}, A., {Halpern}, M.,
  {Hasselfield}, M., {Hern{\'a}ndez-Monteagudo}, C., {Hilton}, G.~C., {Hilton},
  M., {Hincks}, A.~D., {Hlozek}, R., {Huffenberger}, K.~M., {Hughes}, D.~H.,
  {Hughes}, J.~P., {Infante}, L., {Irwin}, K.~D., {Jimenez}, R., {Juin}, J.~B.,
  {Kaul}, M., {Klein}, J., {Kosowsky}, A., {Lau}, J.~M., {Limon}, M., {Lin},
  Y.-T., {Lupton}, R.~H., {Marriage}, T.~A., {Marsden}, D., {Martocci}, K.,
  {Mauskopf}, P., {Menanteau}, F., {Moodley}, K., {Moseley}, H., {Netterfield},
  C.~B., {Niemack}, M.~D., {Nolta}, M.~R., {Page}, L.~A., {Parker}, L.,
  {Partridge}, B., {Quintana}, H., {Reid}, B., {Sehgal}, N., {Sievers}, J.,
  {Spergel}, D.~N., {Staggs}, S.~T., {Swetz}, D.~S., {Switzer}, E.~R.,
  {Thornton}, R., {Trac}, H., {Tucker}, C., {Verde}, L., {Warne}, R., {Wilson},
  G., {Wollack}, E., \& {Zhao}, Y. 2010, \apj, 722, 1148

\bibitem[{{Gnedin} \& {Jaffe}(2001)}]{gnedin01}
{Gnedin}, N.~Y. \& {Jaffe}, A.~H. 2001, \apj, 551, 3

\bibitem[{{Iliev} {et~al.}(2014){Iliev}, {Mellema}, {Ahn}, {Shapiro}, {Mao}, \&
  {Pen}}]{iliev14}
{Iliev}, I.~T., {Mellema}, G., {Ahn}, K., {Shapiro}, P.~R., {Mao}, Y., \&
  {Pen}, U.-L. 2014, \mnras, 439, 725

\bibitem[{{Iliev} {et~al.}(2007){Iliev}, {Pen}, {Bond}, {Mellema}, \&
  {Shapiro}}]{iliev07}
{Iliev}, I.~T., {Pen}, U.-L., {Bond}, J.~R., {Mellema}, G., \& {Shapiro}, P.~R.
  2007, \apj, 660, 933

\bibitem[{{Jeli{\'c}} {et~al.}(2010){Jeli{\'c}}, {Zaroubi}, {Aghanim},
  {Douspis}, {Koopmans}, {Langer}, {Mellema}, {Tashiro}, \& {Thomas}}]{jelic10}
{Jeli{\'c}}, V., {Zaroubi}, S., {Aghanim}, N., {Douspis}, M., {Koopmans},
  L.~V.~E., {Langer}, M., {Mellema}, G., {Tashiro}, H., \& {Thomas}, R.~M.
  2010, \mnras, 402, 2279

\bibitem[{{Jensen} {et~al.}(2014){Jensen}, {Hayes}, {Iliev}, {Laursen},
  {Mellema}, \& {Zackrisson}}]{jensen14}
{Jensen}, H., {Hayes}, M., {Iliev}, I., {Laursen}, P., {Mellema}, G., \&
  {Zackrisson}, E. 2014, ArXiv e-prints

\bibitem[{{Kashlinsky}(2005)}]{kashlinsky05}
{Kashlinsky}, A. 2005, \physrep, 409, 361

\bibitem[{{Kashlinsky} {et~al.}(2012){Kashlinsky}, {Arendt}, {Ashby}, {Fazio},
  {Mather}, \& {Moseley}}]{kashlinsky12}
{Kashlinsky}, A., {Arendt}, R.~G., {Ashby}, M.~L.~N., {Fazio}, G.~G., {Mather},
  J., \& {Moseley}, S.~H. 2012, \apj, 753, 63

\bibitem[{{Lee}(2009)}]{lee09}
{Lee}, K. 2009, ArXiv e-prints

\bibitem[{{Lidz} {et~al.}(2009){Lidz}, {Zahn}, {Furlanetto}, {McQuinn},
  {Hernquist}, \& {Zaldarriaga}}]{lidz09}
{Lidz}, A., {Zahn}, O., {Furlanetto}, S.~R., {McQuinn}, M., {Hernquist}, L., \&
  {Zaldarriaga}, M. 2009, \apj, 690, 252

\bibitem[{{Magliocchetti} {et~al.}(2003){Magliocchetti}, {Salvaterra}, \&
  {Ferrara}}]{magliocchetti03}
{Magliocchetti}, M., {Salvaterra}, R., \& {Ferrara}, A. 2003, \mnras, 342, L25

\bibitem[{{McQuinn} {et~al.}(2005){McQuinn}, {Furlanetto}, {Hernquist}, {Zahn},
  \& {Zaldarriaga}}]{mcquinn05}
{McQuinn}, M., {Furlanetto}, S.~R., {Hernquist}, L., {Zahn}, O., \&
  {Zaldarriaga}, M. 2005, \apj, 630, 643

\bibitem[{{McQuinn} {et~al.}(2007){McQuinn}, {Hernquist}, {Zaldarriaga}, \&
  {Dutta}}]{mcquinn07}
{McQuinn}, M., {Hernquist}, L., {Zaldarriaga}, M., \& {Dutta}, S. 2007, \mnras,
  381, 75

\bibitem[{{Mellema} {et~al.}(2013){Mellema}, {Koopmans}, {Abdalla}, {Bernardi},
  {Ciardi}, {Daiboo}, {de Bruyn}, {Datta}, {Falcke}, {Ferrara}, {Iliev},
  {Iocco}, {Jeli{\'c}}, {Jensen}, {Joseph}, {Labroupoulos}, {Meiksin},
  {Mesinger}, {Offringa}, {Pandey}, {Pritchard}, {Santos}, {Schwarz},
  {Semelin}, {Vedantham}, {Yatawatta}, \& {Zaroubi}}]{mellema13}
{Mellema}, G., {Koopmans}, L.~V.~E., {Abdalla}, F.~A., {Bernardi}, G.,
  {Ciardi}, B., {Daiboo}, S., {de Bruyn}, A.~G., {Datta}, K.~K., {Falcke}, H.,
  {Ferrara}, A., {Iliev}, I.~T., {Iocco}, F., {Jeli{\'c}}, V., {Jensen}, H.,
  {Joseph}, R., {Labroupoulos}, P., {Meiksin}, A., {Mesinger}, A., {Offringa},
  A.~R., {Pandey}, V.~N., {Pritchard}, J.~R., {Santos}, M.~G., {Schwarz},
  D.~J., {Semelin}, B., {Vedantham}, H., {Yatawatta}, S., \& {Zaroubi}, S.
  2013, Experimental Astronomy, 36, 235

\bibitem[{{Mesinger} {et~al.}(2012){Mesinger}, {McQuinn}, \&
  {Spergel}}]{mesinger12}
{Mesinger}, A., {McQuinn}, M., \& {Spergel}, D.~N. 2012, \mnras, 422, 1403

\bibitem[{{Ostriker} \& {Vishniac}(1986)}]{ostriker86}
{Ostriker}, J.~P. \& {Vishniac}, E.~T. 1986, \apjl, 306, L51

\bibitem[{{Ouchi} {et~al.}(2010){Ouchi}, {Shimasaku}, {Furusawa}, {Saito},
  {Yoshida}, {Akiyama}, {Ono}, {Yamada}, {Ota}, {Kashikawa}, {Iye}, {Kodama},
  {Okamura}, {Simpson}, \& {Yoshida}}]{ouchi10}
{Ouchi}, M., {Shimasaku}, K., {Furusawa}, H., {Saito}, T., {Yoshida}, M.,
  {Akiyama}, M., {Ono}, Y., {Yamada}, T., {Ota}, K., {Kashikawa}, N., {Iye},
  M., {Kodama}, T., {Okamura}, S., {Simpson}, C., \& {Yoshida}, M. 2010, \apj,
  723, 869

\bibitem[{{Salvaterra} {et~al.}(2005){Salvaterra}, {Ciardi}, {Ferrara}, \&
  {Baccigalupi}}]{salvaterra05}
{Salvaterra}, R., {Ciardi}, B., {Ferrara}, A., \& {Baccigalupi}, C. 2005,
  \mnras, 360, 1063

\bibitem[{{Salvaterra} \& {Ferrara}(2003)}]{salvaterra03}
{Salvaterra}, R. \& {Ferrara}, A. 2003, \mnras, 339, 973

\bibitem[{{Santos} {et~al.}(2003){Santos}, {Cooray}, {Haiman}, {Knox}, \&
  {Ma}}]{santos03}
{Santos}, M.~G., {Cooray}, A., {Haiman}, Z., {Knox}, L., \& {Ma}, C.-P. 2003,
  \apj, 598, 756

\bibitem[{{Santos} {et~al.}(2002){Santos}, {Bromm}, \&
  {Kamionkowski}}]{santos02}
{Santos}, M.~R., {Bromm}, V., \& {Kamionkowski}, M. 2002, \mnras, 336, 1082

\bibitem[{{Shirokoff} {et~al.}(2011){Shirokoff}, {Reichardt}, {Shaw}, {Millea},
  {Ade}, {Aird}, {Benson}, {Bleem}, {Carlstrom}, {Chang}, {Cho}, {Crawford},
  {Crites}, {de Haan}, {Dobbs}, {Dudley}, {George}, {Halverson}, {Holder},
  {Holzapfel}, {Hrubes}, {Joy}, {Keisler}, {Knox}, {Lee}, {Leitch}, {Lueker},
  {Luong-Van}, {McMahon}, {Mehl}, {Meyer}, {Mohr}, {Montroy}, {Padin},
  {Plagge}, {Pryke}, {Ruhl}, {Schaffer}, {Spieler}, {Staniszewski}, {Stark},
  {Story}, {Vanderlinde}, {Vieira}, {Williamson}, \& {Zahn}}]{shirokoff11}
{Shirokoff}, E., {Reichardt}, C.~L., {Shaw}, L., {Millea}, M., {Ade}, P.~A.~R.,
  {Aird}, K.~A., {Benson}, B.~A., {Bleem}, L.~E., {Carlstrom}, J.~E., {Chang},
  C.~L., {Cho}, H.~M., {Crawford}, T.~M., {Crites}, A.~T., {de Haan}, T.,
  {Dobbs}, M.~A., {Dudley}, J., {George}, E.~M., {Halverson}, N.~W., {Holder},
  G.~P., {Holzapfel}, W.~L., {Hrubes}, J.~D., {Joy}, M., {Keisler}, R., {Knox},
  L., {Lee}, A.~T., {Leitch}, E.~M., {Lueker}, M., {Luong-Van}, D., {McMahon},
  J.~J., {Mehl}, J., {Meyer}, S.~S., {Mohr}, J.~J., {Montroy}, T.~E., {Padin},
  S., {Plagge}, T., {Pryke}, C., {Ruhl}, J.~E., {Schaffer}, K.~K., {Spieler},
  H.~G., {Staniszewski}, Z., {Stark}, A.~A., {Story}, K., {Vanderlinde}, K.,
  {Vieira}, J.~D., {Williamson}, R., \& {Zahn}, O. 2011, \apj, 736, 61

\bibitem[{{Slosar} {et~al.}(2007){Slosar}, {Cooray}, \& {Silk}}]{slosar07}
{Slosar}, A., {Cooray}, A., \& {Silk}, J.~I. 2007, \mnras, 377, 168

\bibitem[{{Sunyaev} \& {Zeldovich}(1970)}]{sunyaev70}
{Sunyaev}, R.~A. \& {Zeldovich}, Y.~B. 1970, \apss, 7, 3

\bibitem[{{Tashiro} {et~al.}(2008){Tashiro}, {Aghanim}, {Langer}, {Douspis}, \&
  {Zaroubi}}]{tashiro08}
{Tashiro}, H., {Aghanim}, N., {Langer}, M., {Douspis}, M., \& {Zaroubi}, S.
  2008, \mnras, 389, 469

\bibitem[{{Tashiro} {et~al.}(2010){Tashiro}, {Aghanim}, {Langer}, {Douspis},
  {Zaroubi}, \& {Jelic}}]{tashiro10}
{Tashiro}, H., {Aghanim}, N., {Langer}, M., {Douspis}, M., {Zaroubi}, S., \&
  {Jelic}, V. 2010, \mnras, 402, 2617

\bibitem[{{Tashiro} {et~al.}(2011){Tashiro}, {Aghanim}, {Langer}, {Douspis},
  {Zaroubi}, \& {Jeli{\'c}}}]{tashiro11}
{Tashiro}, H., {Aghanim}, N., {Langer}, M., {Douspis}, M., {Zaroubi}, S., \&
  {Jeli{\'c}}, V. 2011, \mnras, 414, 3424

\bibitem[{{Vishniac}(1987)}]{vishniac87}
{Vishniac}, E.~T. 1987, \apj, 322, 597

\bibitem[{{Wiersma} {et~al.}(2013){Wiersma}, {Ciardi}, {Thomas}, {Harker},
  {Zaroubi}, {Bernardi}, {Brentjens}, {de Bruyn}, {Daiboo}, {Jelic}, {Kazemi},
  {Koopmans}, {Labropoulos}, {Martinez}, {Mellema}, {Offringa}, {Pandey},
  {Schaye}, {Veligatla}, {Vedantham}, \& {Yatawatta}}]{wiersma13}
{Wiersma}, R.~P.~C., {Ciardi}, B., {Thomas}, R.~M., {Harker}, G.~J.~A.,
  {Zaroubi}, S., {Bernardi}, G., {Brentjens}, M., {de Bruyn}, A.~G., {Daiboo},
  S., {Jelic}, V., {Kazemi}, S., {Koopmans}, L.~V.~E., {Labropoulos}, P.,
  {Martinez}, O., {Mellema}, G., {Offringa}, A., {Pandey}, V.~N., {Schaye}, J.,
  {Veligatla}, V., {Vedantham}, H., \& {Yatawatta}, S. 2013, \mnras, 432, 2615

\bibitem[{{Yue} {et~al.}(2013){Yue}, {Ferrara}, {Salvaterra}, \&
  {Chen}}]{yue13}
{Yue}, B., {Ferrara}, A., {Salvaterra}, R., \& {Chen}, X. 2013, \mnras, 431,
  383

\bibitem[{{Zahn} {et~al.}(2012){Zahn}, {Reichardt}, {Shaw}, {Lidz}, {Aird},
  {Benson}, {Bleem}, {Carlstrom}, {Chang}, {Cho}, {Crawford}, {Crites}, {de
  Haan}, {Dobbs}, {Dor{\'e}}, {Dudley}, {George}, {Halverson}, {Holder},
  {Holzapfel}, {Hoover}, {Hou}, {Hrubes}, {Joy}, {Keisler}, {Knox}, {Lee},
  {Leitch}, {Lueker}, {Luong-Van}, {McMahon}, {Mehl}, {Meyer}, {Millea},
  {Mohr}, {Montroy}, {Natoli}, {Padin}, {Plagge}, {Pryke}, {Ruhl}, {Schaffer},
  {Shirokoff}, {Spieler}, {Staniszewski}, {Stark}, {Story}, {van Engelen},
  {Vanderlinde}, {Vieira}, \& {Williamson}}]{zahn12}
{Zahn}, O., {Reichardt}, C.~L., {Shaw}, L., {Lidz}, A., {Aird}, K.~A.,
  {Benson}, B.~A., {Bleem}, L.~E., {Carlstrom}, J.~E., {Chang}, C.~L., {Cho},
  H.~M., {Crawford}, T.~M., {Crites}, A.~T., {de Haan}, T., {Dobbs}, M.~A.,
  {Dor{\'e}}, O., {Dudley}, J., {George}, E.~M., {Halverson}, N.~W., {Holder},
  G.~P., {Holzapfel}, W.~L., {Hoover}, S., {Hou}, Z., {Hrubes}, J.~D., {Joy},
  M., {Keisler}, R., {Knox}, L., {Lee}, A.~T., {Leitch}, E.~M., {Lueker}, M.,
  {Luong-Van}, D., {McMahon}, J.~J., {Mehl}, J., {Meyer}, S.~S., {Millea}, M.,
  {Mohr}, J.~J., {Montroy}, T.~E., {Natoli}, T., {Padin}, S., {Plagge}, T.,
  {Pryke}, C., {Ruhl}, J.~E., {Schaffer}, K.~K., {Shirokoff}, E., {Spieler},
  H.~G., {Staniszewski}, Z., {Stark}, A.~A., {Story}, K., {van Engelen}, A.,
  {Vanderlinde}, K., {Vieira}, J.~D., \& {Williamson}, R. 2012, \apj, 756, 65

\bibitem[{{Zahn} {et~al.}(2005){Zahn}, {Zaldarriaga}, {Hernquist}, \&
  {McQuinn}}]{zahn05}
{Zahn}, O., {Zaldarriaga}, M., {Hernquist}, L., \& {McQuinn}, M. 2005, \apj,
  630, 657

\bibitem[{{Zeldovich} \& {Sunyaev}(1969)}]{zeldovich69}
{Zeldovich}, Y.~B. \& {Sunyaev}, R.~A. 1969, \apss, 4, 301

\bibitem[{{Zhang} {et~al.}(2004){Zhang}, {Pen}, \& {Trac}}]{zhang04}
{Zhang}, P., {Pen}, U.-L., \& {Trac}, H. 2004, \mnras, 347, 1224

\end{thebibliography}

\end{document}